# Pion Production in Rare Isotope Collisions

M.B. Tsang[1,2], J. Estee[1,2], H. Setiawan[1,2], W.G. Lynch[1,2], J. Barney[1,2], M.B. Chen[2], G. Cerizza[1], P. Danielewicz[1,2], J. Hong[1,2], P. Morfouace[1], R. Shane[1], S. Tangwancharoen[1,2], K. Zhu[1,2], T. Isobe[3], M. Kurata-Nishimura[3], J. Lukasik[4], T. Murakami[5], Z. Chajecki[6], and the SπRIT collaboration

[1]*National Superconducting Cyclotron Laboratory, Michigan State University, East Lansing, Michigan 48824 USA*
[2]*Department of Physics and Astronomy, Michigan State University, East Lansing, Michigan 48824 USA*
[3]*RIKEN Nishina Center, Hirosawa 2-1, Wako, Saitama 351-0198, Japan*
[4]*Institute of Nuclear Physics PAN, Krakow, Poland*
[5]*Department of Physics, Kyoto University, Kita-shirakawa, Kyoto 606-8502, Japan*[5]
[6]*Department of Physics, Western Michigan University, Kalamazoo, Michigan 49008, USA*

**Abstract**

Pion energy spectra are presented for central collisions of neutron-rich $^{132}$Sn+$^{124}$Sn and neutron-deficient $^{108}$Sn+$^{112}$Sn systems using simulations with Boltzmann-Uehling-Uhlenbeck transport model. These calculations, which incorporate isospin-dependent mean field potentials for relevant baryons and mesons, display a sensitivity to the pion spectra that could allow significant constraints on the density dependence of the symmetry energy and its mean field potential at supra-saturation densities. The predicted sensitivity increases with the isospin asymmetry of the total system and decreases with incident energy.

## I.    Introduction

Understanding the nature of dense matter constitutes a significant scientific objective for both nuclear physics and astrophysics [1]. In astrophysics, measurements using ground and satellite based observatories have provided values for the possible masses and radii of neutron stars, raising questions about the nuclear force that support these stars against gravitational collapse into black holes [2]. Clearly, it is incumbent upon nuclear physics to provide laboratory constraints on properties of dense nucleonic matter where such are feasible. Both astronomical observations and nuclear experiments are being designed to address these questions, putting the goal of a quantitative understanding of dense matter within scientific reach [2-6].

This goal requires constraining the nuclear Equation of State (EoS) at densities above and below the saturation density ($\rho_0 \approx 0.16$ nucleons/fm$^3$) that characterizes matter at the centers of atomic nuclei [2-10]. For low temperature systems, such as neutron stars, the EoS, (Eq. 1), can be succinctly given in terms of the energy per nucleon $\varepsilon$ of the matter, the density $\rho$, and the asymmetry $\delta$ of the nuclear system as follows:

$$\varepsilon(\rho, \delta) = \varepsilon(\rho, \delta = 0) + S(\rho)\delta^2 \text{ where } \delta = \frac{\rho_n - \rho_p}{\rho}. \tag{1}$$

Here, $\rho_n$, $\rho_p$ and $\rho$ are neutron, proton and total densities for the system. The second term on the RHS, is known as the symmetry energy. Any system that has very different neutron and proton densities will have $\delta \neq 0$ and be influenced significantly by the symmetry energy and its mean field potential. Constraining the properties of matter within neutron stars requires significant experimental constraints on the symmetry energy at both sub-saturation and supra-saturation densities [3]. For example, analyses on the properties of neutron stars suggest that constraints on the symmetry energy at $\rho/\rho_0 \approx$ 1-2 are relevant to predictions of neutron star radii [2,8]. At such densities the EoS may depend strongly on three-nucleon forces [9,10].

Heavy-ion collisions can be used to momentarily create and study nuclear systems at different density values. The densities to be probed can be varied by changing the incident energy of the beam or the impact parameters of the collisions. Even when the most asymmetric projectiles available at current radioactive ion beam facilities are employed, however, the symmetry energy of available laboratory systems typically contributes less than 20% of the total energy per nucleon in Eq. (1). To maximize the sensitivity to the symmetry energy, one must therefore select observables that minimize the sensitivity to other effects. This can be achieved by comparing the relative emission of members of isospin multiplets, e.g. $\pi^-$ vs. $\pi^+$, *n vs. p, t vs. $^3$He, etc.*, that experience symmetry forces of opposite sign. Another strategy is to compare measurements between pairs of reactions with the same total charge, but very different isospin asymmetries. This choice keeps the Coulomb effects roughly constant while enhancing the effects of the symmetry energy. Such strategies have been used successfully to extract the constraints of the density dependence of the symmetry energy at sub-saturation densities $\rho \leq \rho_0 \approx 0.16$ nucleons/fm$^3$ [10,11,12].

The same strategies are applied in this paper to probe the symmetry energy at $\rho \geq \rho_0 \approx 0.16$ nucleons/fm$^3$, where the uncertainties are much larger. In this work, we study charged pion production in the collisions of the neutron-rich system $^{132}$Sn+$^{124}$Sn and the neutron-deficient system of $^{108}$Sn+$^{112}$Sn. To search for suitable observables and assist in the design of experiments and the data analysis, we simulate heavy-ion collisions with a transport model that solves the Boltzmann-Uehling-Uhlenbeck equation using test particles. These calculations predict that ratios of $\pi^-/\pi^+$ energy spectra measured in Sn+Sn collisions of 300 MeV per nucleon are very sensitive to the symmetry energy at and below twice the saturation density of nuclear matter. In addition, we propose other observables constructed with pion spectra besides the $\pi^-/\pi^+$ ratios, which also display significant sensitivities to the symmetry energy at high density.

We note that pion production has been recently explored within variants of different transport models such as the Boltzman-Uehling-Uhlenbeck (BUU) equation [13,14], Quantum Molecular Dynamics approaches (QMD) [15, 16] and the Anti-Symmetrized Molecular Dynamics (AMD) model [17]. Most of these calculations have computed the ratio of the total yield of negative pions to the total yield of positive pions. This results in small and sometimes inconsistent sensitivities to the density dependence of the symmetry energy [16,18]. Such inconsistencies may stem, in part, from the neglect of the mean field potentials for the Δ resonances or for the produced pions or both, since these potentials can have an effect on the pion production thresholds within matter [13, 14]. In the present work, energy spectra for pions are calculated using the version of pBUU described in ref. [13], which includes mean field potentials for the relevant baryons and mesons (pions) and optimized to describe the production of pions and nucleon flows

## II. Laboratory constraints at supra-saturation densities

The equation of state for symmetric matter, $E_0(\rho, \delta=0)$, and its pressure have been constrained at densities ranging from saturation density to five times saturation density by measurements of isoscalar collective vibrations, collective flow and kaon production in energetic nucleus-nucleus collisions [7,19]. The density dependence of the symmetry energy, $S(\rho)\delta^2$, is not well determined and therefore contributes a large uncertainty to the EoS for neutron star matter [2,3,8]. While many nuclear structure and reaction observables can constrain the EoS for neutron-rich matter at sub-saturation densities [3,10, 20], laboratory constraints at supra-saturation densities, where the uncertainties are greatest, can only be provided via measurements of heavy-ion collisions.

In the peripheral collisions of projectile and target nuclei of different asymmetry, the difference in the symmetry energies causes isospin dependent diffusion that brings the two nuclei closer to isospin equilibrium. Comparisons of calculated and measured values of this isospin diffusion have provided constraints on the density dependence of the symmetry energy at $\rho \approx 0.45\ \rho_0$ [12].

In more central collisions, differences in the emission of neutrons and protons from heavy-ion collisions result from differences in the (Coulomb and symmetry) mean field potentials that accelerate these particles. Recent measurements of the differences in the neutron and proton emission rates can allow constraints on the density and momentum dependencies of the symmetry mean field potential at $0.3\rho_0 \leq \rho \leq 2\ \rho_0$ [21,22]. Here, the mean field potentials either increase (for neutrons) or decrease (for protons) the relevant pressures within the expanding nuclear system. However, each of these mean field effects can be expected to diminish as the temperature of the system is increased [23,24]. Thus at incident energy much above E/A=100 MeV where nuclear systems above normal density can be created, many nucleonic isospin observables become weaker and somewhat more difficult to use.

To gain additional sensitivity to the EoS and symmetry energy at high densities, $\pi^-/\pi^+$ yield and spectral ratios have been proposed to probe the symmetry energy at supra-normal density [13,17,25]. Within transport theory calculations, subthreshold pions are produced at the highest density via the excitation and decay of Δ resonances. Despite this agreement about the basic pion production mechanism, there are discrepancies in published transport theoretical predictions for $\pi^-/\pi^+$ pion ratios [13,17,18,26]. It is expected that new data will become available from experiments aimed at probing the symmetry energy at $\rho \approx 2\rho_0$ using neutron-rich and neutron deficient rare isotope beams [4,27]. It is, therefore, timely to try to understand the source of such discrepancies.

One source of such discrepancies may lie in the inconsistencies in the mean field potentials for the relevant baryons and mesons [13,14]. Generally, the neutron and proton mean field potentials are specified by the equation of state, which enforces consistency in the nucleonic mean field potentials among calculations. Pions are primarily produced at intermediate energies via Δ resonance production and decay. Different choices regarding the mean field potentials for pions and Δ resonances will influence the thresholds for pion production in transport theories [13,14, 28]. In particular, the choices of symmetry mean field potentials for $\Delta^{++}$, $\Delta^+$, $\Delta^o$, $\Delta^-$ and for $\pi^+$, $\pi^0$ and $\pi^-$ can be important [13,14, 28].

### III. Simulation Details

Most theoretical calculations have focused on the ratio of the net yields of positive and negative pions measured in Ref. [29]. This data set does not provide sufficient information to address some of the contradictory interpretations of theoretical analyses by refs. [15, 16, 18] of the $\pi^-/\pi^+$ total yield ratios for Au+Au collisions. Specifically, the published pion data in Ref. [29] are limited to the net yields of charged pions; the pion energy spectra were not obtained. Thus, the roles of Coulomb effects and the pionic optical potentials [30-32] on the relative production of positive and negative pions cannot be adequately tested by that data, nor can the data sufficiently constrain the theoretical modeling of such effects. As shown below, the sensitivity of $\pi^-/\pi^+$ spectral ratios to Coulomb effects and pion optical model parameters is considerable, and their treatment in the extrapolations of measured pion yields to the experimental thresholds may be correspondingly difficult to control as well.

To avoid the above problems in our simulations, we calculate pion spectra to high statistical precision and strive to unambiguously separate the effects of Coulomb and symmetry potentials by explicitly comparing pairs of reactions using Sn isotopes and also to search for different combinations of observables to enhance the Coulomb (manifested in low energy pions) and symmetry energy effects (manifested in high energy pions) separately. We simulate two reactions, $^{132}$Sn+$^{124}$Sn (neutron rich) and $^{108}$Sn+$^{112}$Sn (neutron deficient) to maximize the symmetry energy effect at a central impact parameter of 3 fm. The two reactions are among those measured by the SπRIT collaboration at the Radioactive Isotope Beam Factory (RIBF) in RIKEN, Japan [4, 27, 33]. Simulations are performed at two energies, E/A=200 and 300 MeV, which are feasible at the RIBF.

The present calculations were performed using the pBUU version of the Boltzmann-Uehling-Uhlenbeck transport model code developed by Danielewicz and Hong in ref. [13]. These calculations utilize a symmetric matter EoS consistent with flow measurements and with an incompressibility constant of K=230 MeV and an isoscalar effective mass of $\frac{m^*}{m} = 0.75$ [7]. The $N\pi$-adjusted mean field parameterization which is optimized to describe available data on pion production is used. The symmetry energy potential term in Equation 1 is parameterized as follow:

$$S(\rho) = S_{kin} \left(\frac{\rho}{\rho_0}\right)^{\frac{2}{3}} + S_{int} \left(\frac{\rho}{\rho_0}\right)^{\gamma} \qquad (2)$$

where the first term on the right is the kinetic term with $S_{kin}$ = 12.3 MeV and the second term with $S_{int}$=20 MeV is the interaction term. A stiff symmetry energy density dependence (such as γ=1.75) predicts that symmetry energy is strongly dependent on density, while a soft dependence (such as γ=0.5) predicts a weaker density dependence proportional to square-root of density, as illustrated by Fig. 1. The "stiffer" symmetry mean field with larger γ has larger symmetry pressure for ρ>ρ$_0$ and expel neutrons more effectively than the "softer" mean field with smaller γ [34]. The opposite is true at ρ<ρ$_0$ [34].

The pBUU code solves the Boltzmann-Uehling-Uhlenbeck equations by the test particle method. The mass of clusters in the current pBUU code are limited to A<4. To gain statistics for pions, whose multiplicities are two orders of magnitude lower than the proton or neutron multiplicities, typically 1000 simulated collisions were performed for each system, each collision containing 3000 test particles. A more detailed description of the formalism and parameters of these calculations is provided in ref. [13].

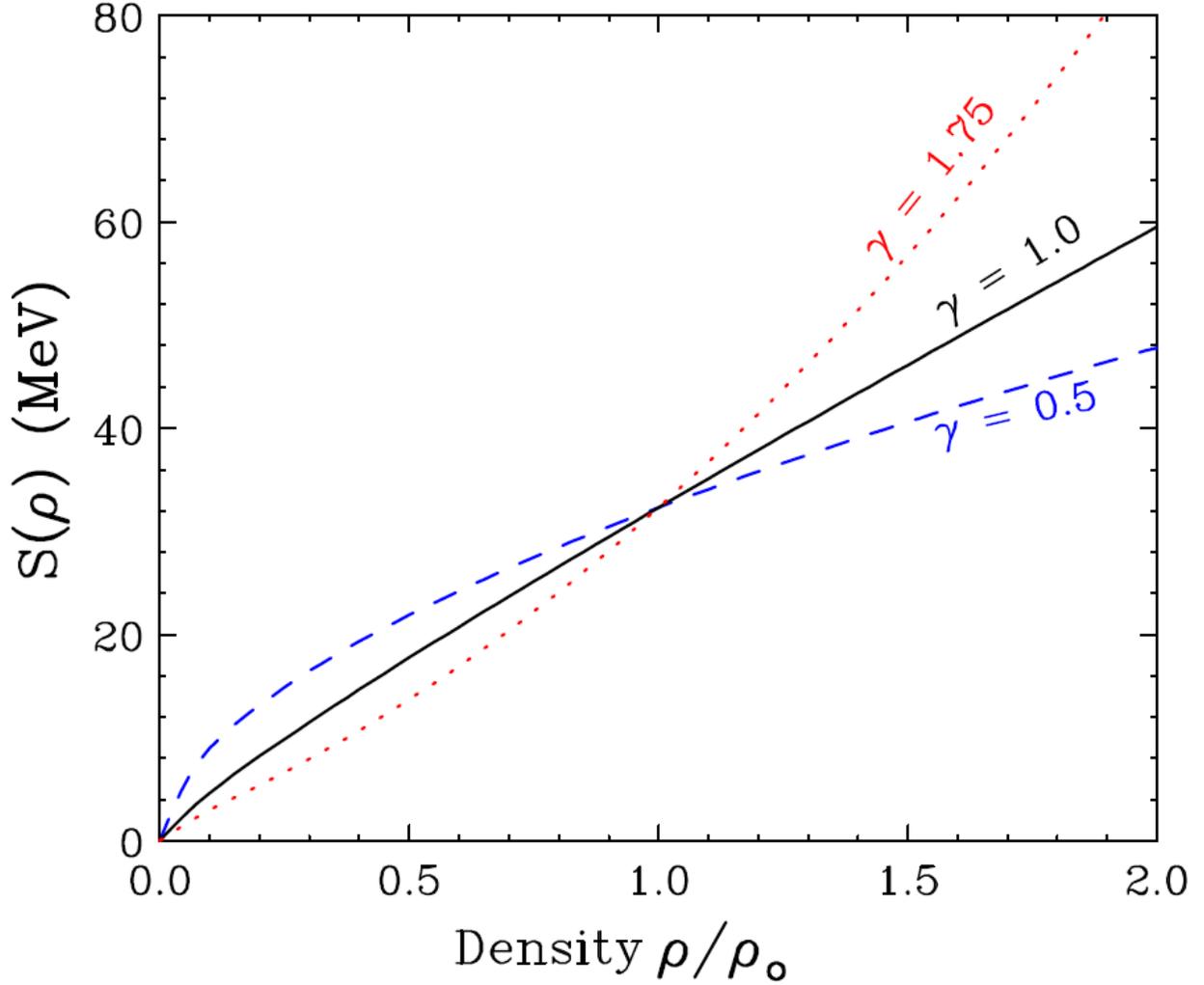

Figure 1: The density dependence of the symmetry energy is shown for three different values of the parameter γ of Eq. (2) that controls the density dependence of the potential energy component of the symmetry energy.

### IV.     Results

At incident energy of E/A=300 MeV, these calculations predict the emission of pions to be a comparatively rare process that occurs with an average multiplicity of the order of unity or less per collision. In central collisions, the projectile nuclei overlap and compress to nearly twice saturation density and subsequently expand and disintegrate. The calculation predicts that most of the final particles in the collision are neutrons or protons, but the production of clusters with A<4, i.e., deuterons, tritons ($^3$H), and $^3$He, is also predicted [13]. Experimental observables constructed from the momenta of these particles reflect a complex interplay between the dynamics of motion in Coulomb, isoscalar and symmetry mean field potentials and that induced by nucleon-nucleon collisions due to the residual interactions. The investigation of this interplay requires more information than one can obtain from the total pion yield ratios as first proposed in ref. [25].

Fig. 2 shows the particle energy spectra produced in central (b=3 fm) $^{132}$Sn+$^{124}$Sn collisions at E/A=300 MeV incident energy with γ=0.5. In addition to the pressure of the symmetry energy, these spectra reflect the forces due to the symmetric matter EoS, the random (thermal) impulses from the residual interactions, as well as any undamped motion along the beam axis which dominates final momenta of unscattered "spectator" nucleons. For this reason, the difference in the radial flow for neutrons and protons, and ratios of neutron and proton energy spectra is predicted to display a complex dependence on the maximum density achieved in a collision and, consequently, on the incident energy. To reduce the trivial contributions from spectator nucleons and to concentrate on the compression induced radial flow that influences the participant nucleons, we select particles from the mid-rapidity source by integrating the differential multiplicities $dM/d\Omega_{cm} dE_{cm}$ over the azimuthal angle and over polar angles $60^0 < \theta_{cm} < 120^0$ in a center of mass system defined with the polar axis along the beam direction as follows:

$$dM/dE_{cm} = \int_{60}^{120} d\theta_{cm} \sin(\theta_{cm}) \int_0^{2\pi} d\vartheta \cdot dM/d\Omega_{cm} dE_{cm}$$

The corresponding energy spectra dM/dE$_{cm}$ of n, p, t, $^3$He, $\pi^-$ and $\pi^+$ are shown in Fig. 2. Nucleons have the largest integrated yields and for a neutron-rich system, the yields of n, t and $\pi^-$ are larger than their respective mirror isospin counterparts p, $^3$He and $\pi^+$. In this figure, the n and p spectra have been multiplied by a factor of 10 to clearly separate them from the spectra of A=3 particles.

The pions are produced early in the collisions from the decay of delta resonances formed via nucleon-nucleon collisions in the nuclear medium. Detailed accounting for the decay and formation matrix elements shows that nn collisions and pp collisions are primarily responsible for $\pi^-$ and $\pi^+$ production, respectively. Measurements of negative and positive pions therefore reflect the neutron and proton abundance in the high-density overlap region [13, 17] and in turn, the symmetry energy, which in neutron-rich regions expels neutrons and attracts protons. While some information can be lost by pion rescattering and reabsorption, energetic pions are predicted to retain information about the symmetry energy and its mean field potentials. Unlike the charged particles, especially the heavier fragments, pions are minimally affected by the radial flow and therefore the rapidity cut. Such observation of a weak effect is consistent with the directional effect investigated in Ref. [13]. Since the mid-rapidity gates cut out nearly half of the pions, we do not impose this gate on the pion calculations shown after Fig. 2.

While the $\pi^-$ and $\pi^+$ energy spectra have similar slopes as n and p spectra, their yields are two orders of magnitude smaller. This can be expected from strongly negative value of the Q-value (Q≈-140 MeV) for producing pions in nucleon-nucleon collisions, much more negative than the typical Q-value for the predominant decay mode of emitting a nucleon. The pion production threshold for pp→pnπ$^+$ or nn→npπ$^-$ reactions is ~280 MeV. Neglecting binding energy and Fermi motion effects, Sn+Sn collisions at incident energy of E/A=300 MeV is above pion production threshold but below the threshold at E/A=200 MeV. Accordingly, more collisions successfully produce pions at E/A=300 MeV than at E/A=200 MeV. Moreover as the incident kinetic energy is dissipated with time, the probability of producing a pion progressively decreases.

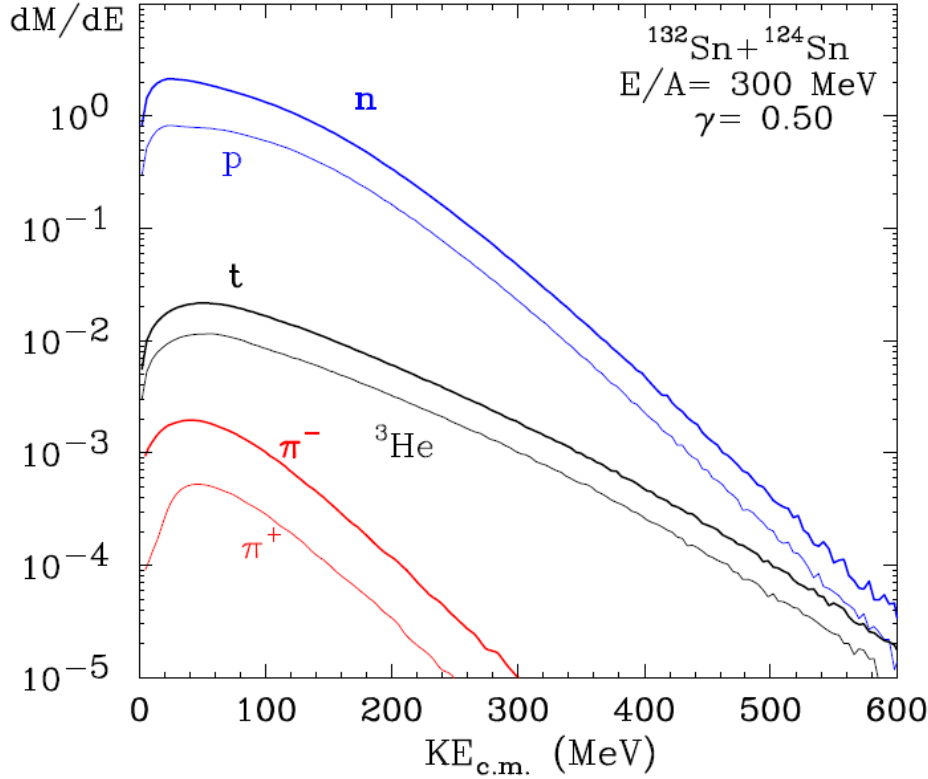

Figure 2: Center of mass energy spectra for neutrons, protons, tritons $^3$He, $\pi^-$ and $\pi^+$ particles emitted in central (b=3fm) $^{132}$Sn+$^{124}$Sn collisions at E/A=300 MeV incident energy. These calculations were performed assuming a mean field corresponding to a soft symmetry energy with γ=0.5 in Eq. (2). For clarity in presentation, the n and p spectra are increased by a multiplicative factor of 10.

Fig. 3 shows the energy spectra of charged pions produced in central (b=3 fm) $^{132}$Sn+$^{124}$Sn collisions at E/A=300 MeV. Results for two different stiffness parameters, γ=0.5 (dashed) and γ=1.75 (solid lines), are shown. As expected for a neutron-rich system, many more negative pions than positive pions are emitted. The Coulomb force plays an important role in the emission of charged pions. It boosts the $\pi^+$ to higher energy resulting in a Coulomb peak at around 40 MeV in the energy spectra. In contrast, the $\pi^-$ spectra reach their maximum values at a lower energy of about 10 MeV. At supranormal densities, the force from symmetry potential increases with γ. Calculations with larger values of γ, i.e. assuming a stiffer symmetry energy, predict neutrons to be ejected more readily from the collision region. Consequently, fewer high energy (KE$_{CM}$ >50 MeV) $\pi^-$ and more high energy $\pi^+$ are emitted in the collision.

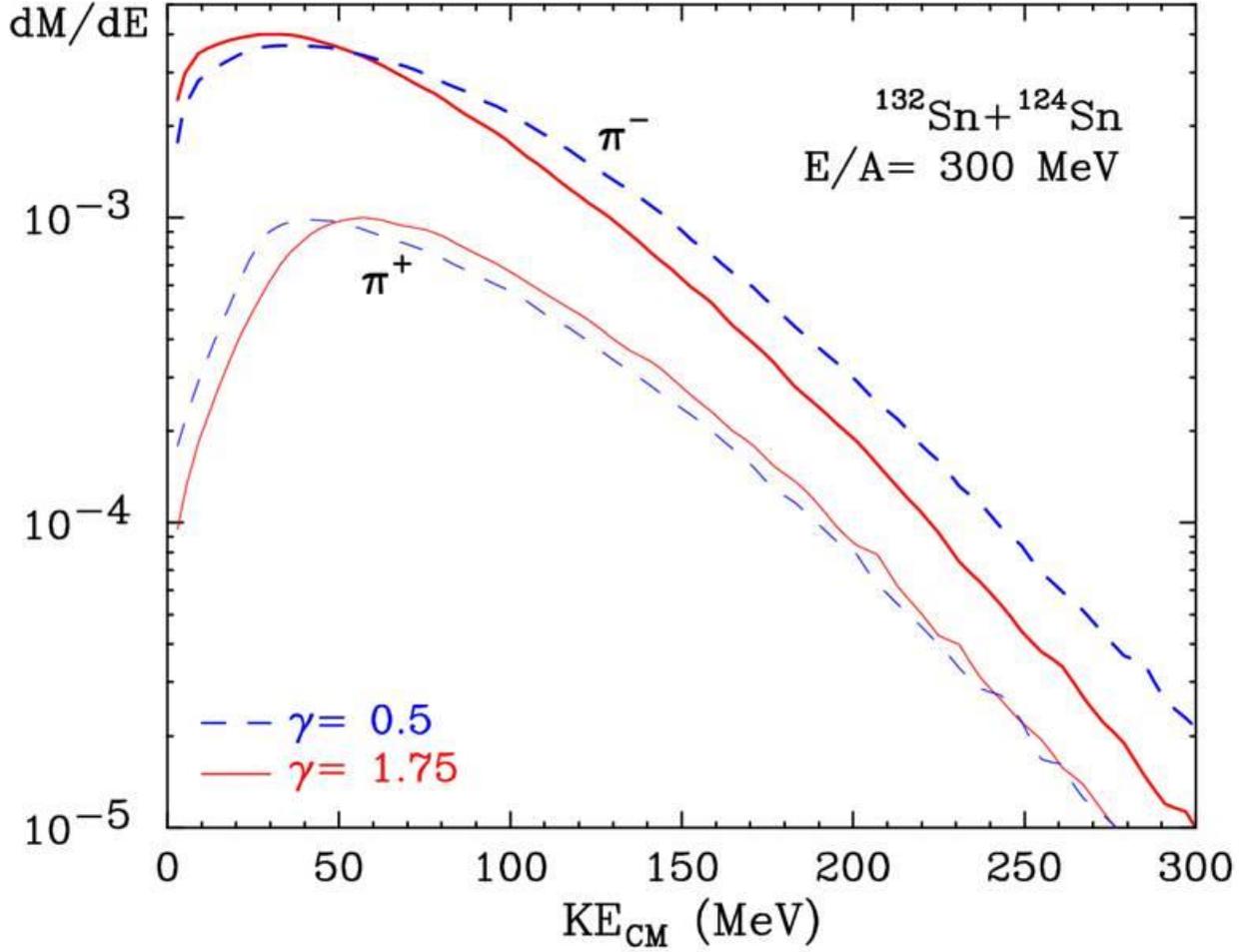

Fig. 3: Pion energy spectra for the $^{132}$Sn+$^{124}$Sn reaction at b=3 fm for two symmetry energy parameters of γ=0.5 (soft) and γ=1.75 (stiff).

To further enhance the sensitivity of the spectra to the symmetry energy, we divide the $\rho^-$ spectra by the $\rho^+$ spectra to obtain the $\pi^-/\pi^+$ spectral ratios shown in Fig. 4 for two reaction systems $^{132}$Sn+$^{124}$Sn (left panels) and $^{108}$Sn+$^{112}$Sn (right panels) at incident energies of E/A=300 MeV (bottom panels) and E/A=200 MeV (top panels) incident energies. Calculations using a soft symmetry energy as input are shown by the blue shaded regions, while corresponding calculations for a stiff symmetry energy are shown as the red shaded regions. Pion ratios of soft and stiff systems differ by a factor of two in the neutron-rich system of $^{132}$Sn+$^{124}$Sn (left panels), but the difference is much smaller in the nearly symmetric system of $^{108}$Sn+$^{112}$Sn (right panels). The ratio for the soft symmetry energy and the neutron rich $^{132}$Sn+$^{124}$Sn system is larger and less dependent on pion energy than the ratio for the stiff symmetry energy. In addition, there is a larger decrease in the spectral ratios going from the $^{132}$Sn+$^{124}$Sn reaction to the $^{108}$Sn+$^{112}$Sn reaction for the soft symmetry energy (γ=0.5) than for the stiff (γ=1.75) symmetry energy. Not shown here are the n/p and t/$^3$He spectral ratios which are smaller and less sensitive to the density dependence of the symmetry energy. The results suggest that pions produced early in the collisions at the higher density neutron rich regions, are heavily influenced by the symmetry energy parameter (γ), predicting that it will be an interesting observable to study symmetry energy.

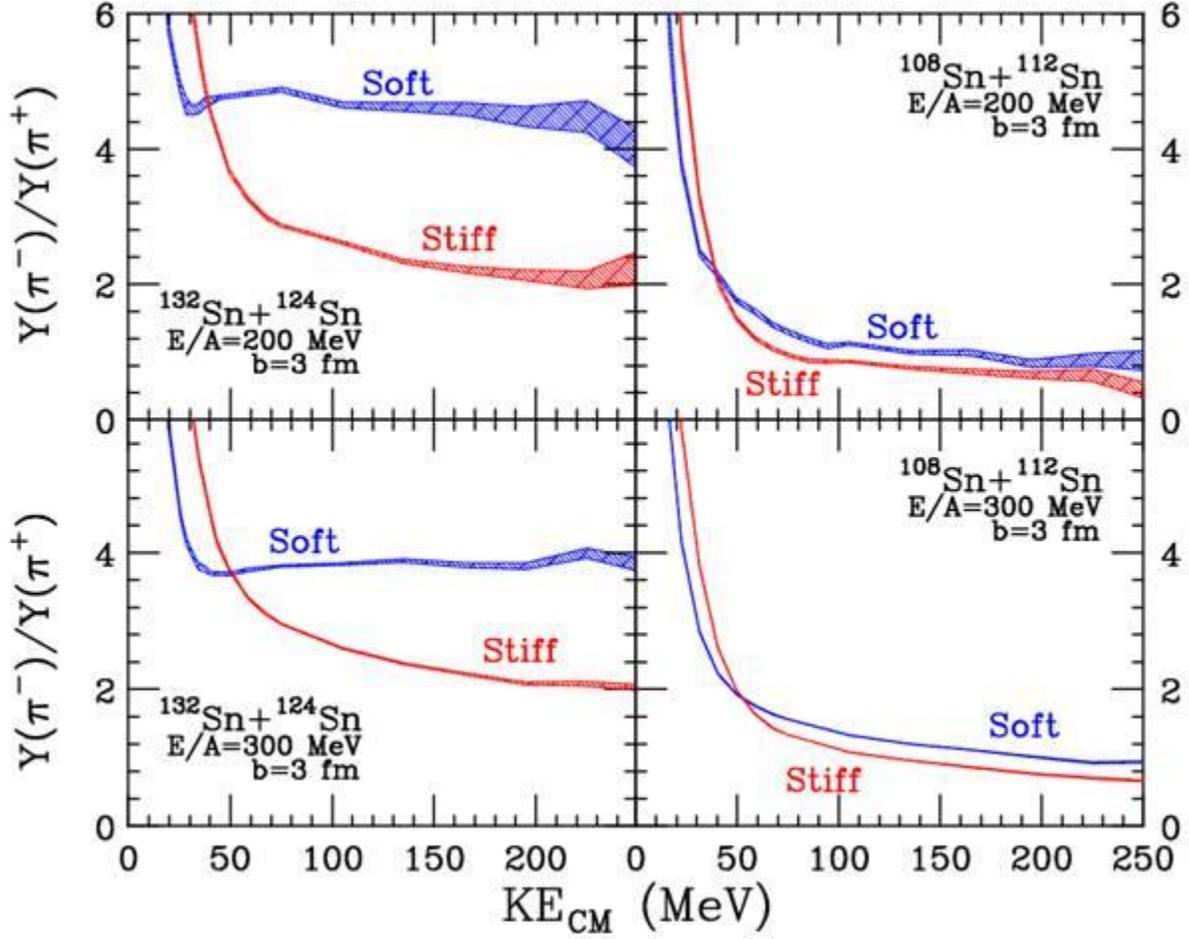

Figure 4: Comparison of $\pi^-/\pi^+$ spectral ratios from central (b=3fm) collisions of $^{132}$Sn+$^{124}$Sn (left panels) and $^{108}$Sn+$^{112}$Sn (right panels) reactions at the incident beam energies of 200 MeV (upper panels) and 300 MeV (lower panels) per nucleon. Calculations for both soft ($\gamma = 0.5$) and stiff ($\gamma=1.75$) symmetry energies are shown.

    In anticipation of results from experiments recently performed with the SπRIT Time Projection Chamber described in Ref. [4], it is worthwhile to consider observables chosen to minimize systematic experimental uncertainties. For example, negative and positive pions are detected in different regions of phase space characterized by different detection efficiencies. Negative pions will be much easier to identify because they are the main particle with negative charge and will be bent by the magnetic field in the opposite direction from that of positively charged fragments such as p, d, t, $^3$He, $^4$He etc. and positive pions. The contaminations and background will be much larger for π$^+$ than for π$^-$ experimentally. To reduce significantly this efficiency difference, one can construct double pion ratios by dividing the $\pi^-/\pi^+$ spectral ratio for the $^{132}$Sn+$^{124}$Sn reaction by the $\pi^-/\pi^+$ spectral ratio for the $^{108}$Sn+$^{112}$Sn reaction. Such double ratios are shown in the left panels of Fig. 5 for the stiff and soft symmetry energies. These ratios show a strong sensitivity to the density dependence of the symmetry energy. Even though the effect is smaller than the single particle ratios in Figure 4, it is not influenced by systematic errors in the detection efficiency to a large extent. The sensitivity to the symmetry energy parameter (γ), increases with

energy of the pions. The influence of the Coulomb force and of pion optical potentials at low kinetic energies is also significantly reduced. We also explore another observable that takes the difference of the single ratios for the two reactions; this is shown in the right panels. This difference shows both a sensitivity at high energy to the symmetry energy, as well as a significant differences at the low energy, where pions are especially sensitive to the Coulomb and the optical potentials. This suggests the plausibility of using the ratio differences to enhance sensitivity to the isospin dependence of the pion optical potential and maybe the Δ symmetry potential [13,14].

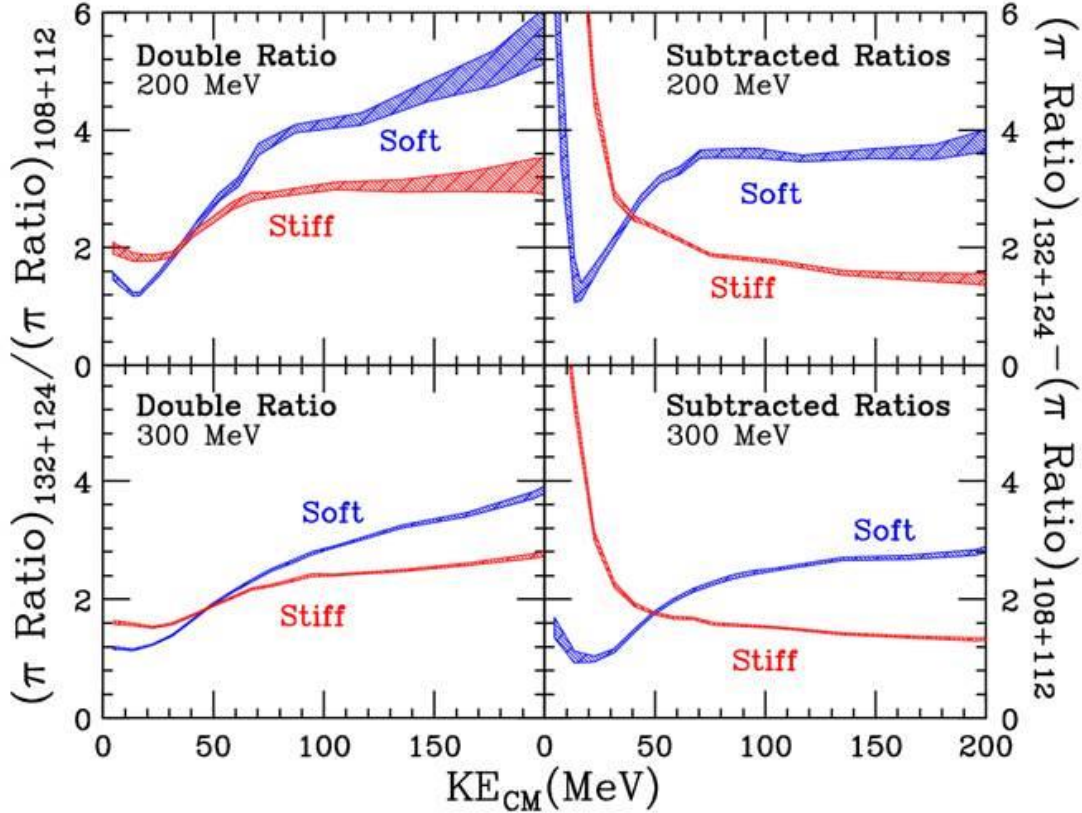

Figure 5: Double ratios (left panels) and difference (right panels) of $\pi^-/\pi^+$ spectral ratios from central (b=3fm) collisions of $^{132}$Sn+$^{124}$Sn and $^{108}$Sn+$^{112}$Sn reactions at the incident beam energies of 200 MeV (upper panels) and 300 MeV (lower panels) per nucleon. Calculations for both soft ($\gamma = 0.5$) and stiff ($\gamma=1.75$) symmetry energies are shown.

To minimize the systematic errors in detecting pions, one can also construct isoscaling ratios by dividing the $\pi^-$ spectra for the $^{132}$Sn+$^{124}$Sn reaction by the $\pi^-$ spectra for the $^{108}$Sn+$^{112}$Sn reaction. Likewise, one can construct an equivalent ratio from the $\pi^+$ spectra. Such isoscaling ratios for $\pi^-$ (left panel) and $\pi^+$ (right panel) are shown in Fig. 6. Due to their negative charges, detecting negative pions and constructing the $\pi^-$ isoscaling ratio is much easier than performing the corresponding measurements and analyses of the $\pi^+$ isoscaling ratio which display a similar but much smaller sensitivity to the symmetry energy.

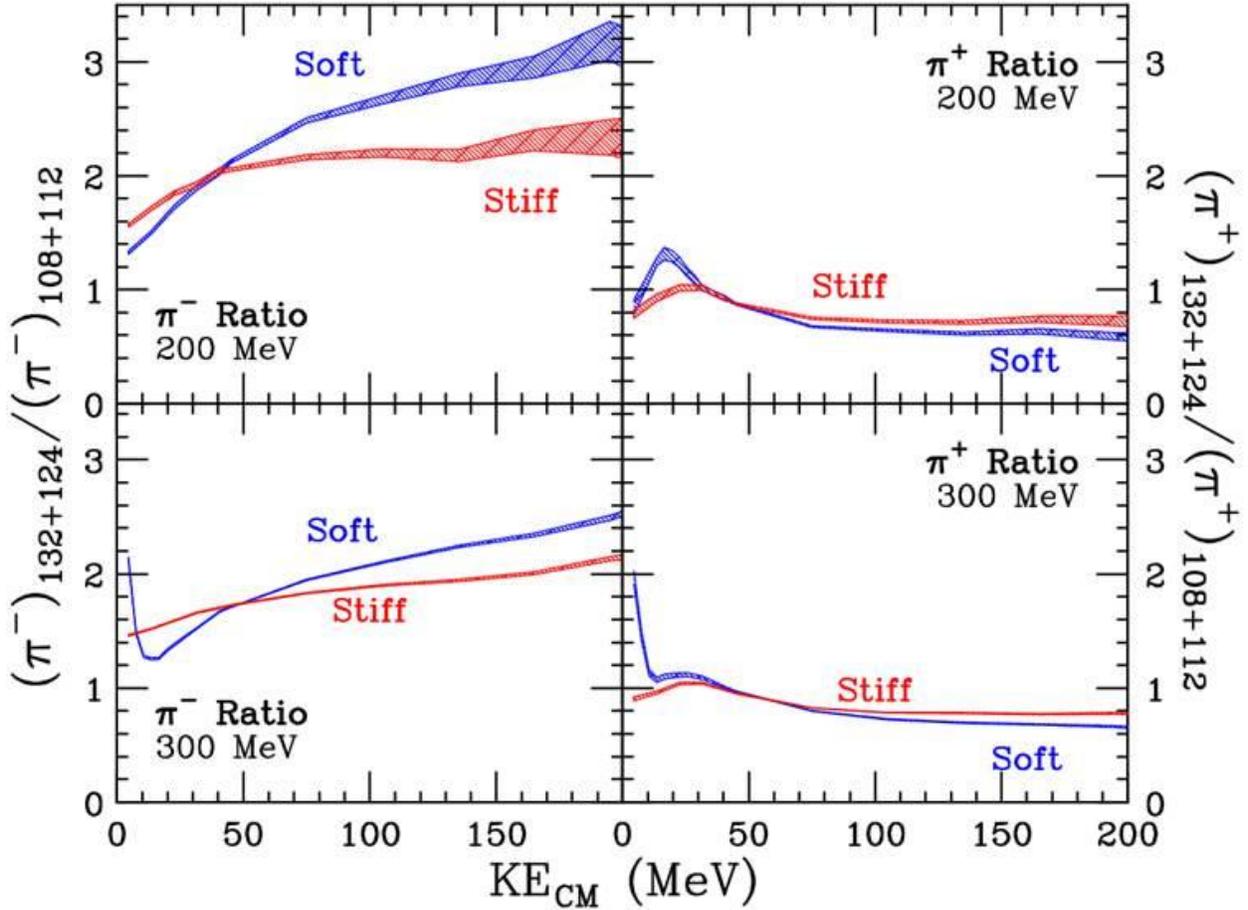

Figure 6: Isoscaling ratios of $\pi^-$ (left panels) and $\pi^+$ (right panels) from central (b=3fm) collisions of $^{132}$Sn+$^{124}$Sn and $^{108}$Sn+$^{112}$Sn reactions at the incident beam energies of 200 MeV (upper panels) and 300 MeV (lower panels) per nucleon. Calculations for both soft ($\gamma = 0.5$) and stiff ($\gamma=1.75$) symmetry energies are shown.

    In general, these isospin signals decrease in sensitivity with increasing beam energy as the thermal or random energies become more comparable and eventually larger than the potential energy differences between the two symmetry energies. One might thus expect larger pion ratios at lower incident energy. The top panels of Figures 4-6 show the single, double, difference and isoscaling ratios at incident energy of E/A=200 MeV. The bottom panels show the corresponding ratios at E/A=300 MeV. It is unmistakable that the effects are larger at the lower energy although the reduced pion yields will make the measurements much more time consuming and suffer from statistics. However, data from different energy and density regions provide an independent test of the models. Furthermore, there is no reason to expect the symmetry energy to depend as a power law on the density. It may be that the value of the power γ that best fits the data at lower energy is different than the best-fit value at higher energies, allowing to discern the true density dependence of the symmetry energy.

**Summary and Future Directions**

In summary, simulations using the pBUU code has been performed for the $^{132}$Sn+$^{124}$Sn (neutron rich) and $^{108}$Sn+$^{112}$Sn (neutron deficient) reactions. By constructing spectra ratios, it is possible to separate the Coulomb effects from the isospin (symmetry energy) effects. We also discuss the pros and cons of constructing singles and double ratios. Taking advantage of its negative charge, one expects that $\pi^-$ should be easier to identify in the presence of a magnetic field in a Time Projection Chamber as $\pi^-$ will be deflected to the opposite side of the beam than are $\pi^+$ and other positive charged particles. We thus propose that measurements of the isoscaling ratio of $\pi^-$, which is shown to be sensitive to the strength of the symmetry energy, should be added to the list of observables. Since such experiments have recently been performed, there are excellent prospects for testing the ideas presented in the paper when the data are analyzed and become available.

### Acknowledgements

This work is supported by the U.S. Department of Energy under Grant Nos. DE-SC0004835, DE-SC0014530, DE-NA0002923 and NSF Grant No. PHY 1565546, PHY 1510971, PHY 1430152. The pBUU simulations were carried out at the Institute for Cyber-Enabled Research (ICER) at MSU.